\documentclass[prb,showpacs,twocolumn,superscriptaddress]{revtex4-1}
\usepackage{amssymb}
\usepackage{graphicx}
\usepackage{amsmath,amssymb}

\begin{document}

\title{Rashba induced Kondo screening of a magnetic impurity in two-dimensional superconductor}
\author{Lin Li}
\affiliation{Department of Physics, Southern University of Science and Technology of China, Shenzhen 518005, China}
\author{Ming-Xin Gao}
\affiliation{Center of Interdisciplinary Studies and Key Laboratory for Magnetism and
Magnetic Materials of the Ministry of Education, Lanzhou University, Lanzhou 730000, China}
\author{Zhen-Hua Wang}
\affiliation{Department of Physics, Southern University of Science and Technology of China, Shenzhen 518005, China}
\affiliation{Beijing Computational Science Research Center, Beijing 100084, China}
\author{Hong-Gang Luo}
\affiliation{Center of Interdisciplinary Studies and Key Laboratory for Magnetism and
Magnetic Materials of the Ministry of Education, Lanzhou University, Lanzhou 730000, China}
\affiliation{Beijing Computational Science Research Center, Beijing 100084, China}
\author{Wei-Qiang Chen}
\affiliation{Department of Physics, Southern University of Science and Technology of China, Shenzhen 518005, China}

\begin{abstract}
We study the Kondo screening of a magnetic impurity in a two-dimensional superconductor with Rashba spin-orbit coupling (SOC). It is found that the Rashba interaction generates a novel Kondo screening channel, in which the local moment is screened by the exchange coupling with conduction electrons in different spin and orbital states. The Kondo temperature associated with this process is determined by the interplay between the Rashba SOC and superconducting energy gap. As a result, the quantum phase transition between the magnetic doublet and Kondo singlet ground states is significantly affected by increasing Rashba SOC in such a system. This result uncovers that the Rashba SOC plays an instructive role and provides a novel screening channel for the Kondo effect, which is expected to be observed in future experiments.
\end{abstract}

\maketitle

\section{Introduction}
The Kondo effect, originating from the screening of magnetic moment by conduction electrons, is one of the well-understood phenomena in many-body physics\cite{Kondo1964,Hewson1993}. In artificial nanostructures, the Kondo effect manifests itself as a zero-bias resonance peak at low temperatures \cite{Madhavan1998,Kouwenhoven1998,Kouwenhoven2000}.
When a magnetic impurity is immersed into a superconductor, due to the exchange scattering with Cooper pairs, the magnetic moment can induce subgap Yu-Shiba-Rusinov(YSR) bound states \cite{Yu1965,Shiba1968,Rusinov1969,Zhu2006}.
The behavior of the YSR bound states can uncover many interesting physics, for example, the interplay between the Kondo screening and superconducting pair-breaking interactions, which leads to two different ground states, namely, the magnetic doublet state and the Kondo singlet state \cite{Clerk1999,Siano2004,Buizert2007,Eduardo2012,Franke2011,Kim2013,Chang2013}.
The quantum phase transition (QPT) between these two ground states takes place at $T^{0}_K/\Delta\sim 1$, which can be characterized by the level-crossing of the YSR bound states ($T^{0}_{K}$ is the normal state Kondo temperature, and $\Delta$ is the superconducting energy gap)\cite{Franke2011,Kim2013,Chang2013,Li2014,Island2017}.
Recently, many works have shown that a superconductor with strong spin-orbit coupling (SOC) can provide many interesting features such as Majoranan zero mode found in Fe or Co chain \cite{Perge2014,Ruby2015,Feldman2016,Peng2015, Sau2015, Schecter2016, Zhang2016, Poyhonen2016}.
This motives us to consider Kondo physics of an adatom on the surface of a superconductor with strong SOC. 
In such systems, the SOC should play an important role in Kondo screening.

In the literature there exists many works to discuss the influence of the SOC on the Kondo effect. If the SOC is present in the magnetic impurity or quantum dot, for single-wall carbon nanotube quantum dot\cite{Kuemmeth2008}, the SOC leads to the fine splitting of the Kondo resonance\cite{Fang2008}, which has been experimentally observed by different experimental groups\cite{Lan2012, Cleuziou2013, Lai2014}. Here we focus on the cases that the SOC is present in the environmental conduction electrons, for example, those electrons on the surface of the three-dimensional topological insulator\cite{Zhang2009}. In the presence of SOC, the spin and orbital angular momentum are not conserved separately. It is natural to represent the conduction electrons in orbital angular momentum basis with a definite $z$ component about the impurity.\cite{Isaev2012, Zarea2012,Isaev2015} Then, the total angular momentum $j_z(=l_z+s_z)$ is a conserved quantity, $l_z$ and $s_z$ are the orbital angular momentum and spin of conduction electrons, respectively.
In these cases, the Kondo resonance could be enhanced by the weak Rashba or Dresselhaus SOC \cite{Isaev2012, Zarea2012}. In particular, for strong SOC, Isaev, Ortiz and Vekhter (IOV) \cite{Isaev2015} even found a new mechanism for the Kondo screening, namely, the impurity spin is screened by purely orbital degree of freedom of the surface electrons in the three-dimensional topological insulator. These works show rich physics involved in the Kondo effect and motivate us to further explore Kondo screening for a magnetic impurity absorbed on a two-dimensional electron gas (2DEG) proximity to a conventional superconductor.

\begin{figure}[tbp]
\includegraphics[clip=true,width=1.0\columnwidth]{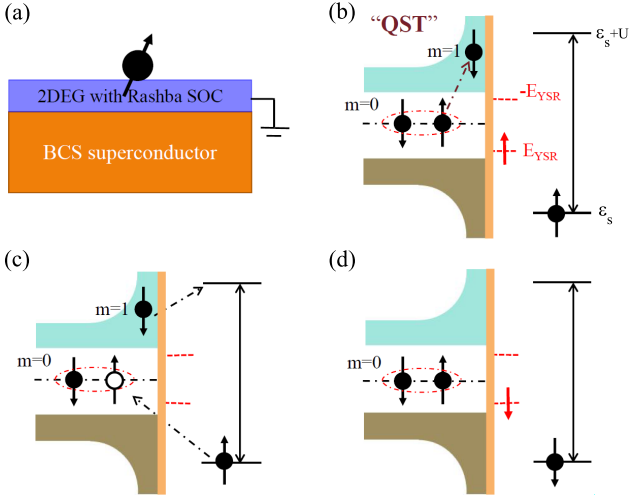}
\caption{(a) Schematic diagram of an Anderson impurity on 2DEG with Rashba spin-orbit coupling (SOC) proximity to BCS superconductor. (b) The magnetic moment induces the Yu-Shiba-Rusinov (YSR) bound states at $E_{YSR}$, which possesses effective moment denoted by the red arrow \cite{Li2016}. The Rashba SOC leads to a total angular momentum conserved QST process, like ($\left\vert 0,\uparrow \right\rangle \rightarrow \left\vert 1,\downarrow\right\rangle $). (c) The spin-up electron on impurity level is replaced by the excited electron ($\left\vert1,\downarrow\right\rangle$) through the virtual states. And the spin-up electron tunnels into the conduction band to neutralize the hole ($\left\vert0,\uparrow\right\rangle$) in superconductor. (d) The local moment of impurity as well as the effective moment of YSR bound states are effectively flipped. The magnetic impurity is screened by the coherent superposition of those processes.}
\label{fig1}
\end{figure}

The setup we study is shown schematically in Fig.\ref{fig1}\,(a).
The Kondo screening originating from pair-breaking exchange interaction between the magnetic impurity and Cooper pairs has been intensively in the literature.\cite{Clerk1999,Siano2004,Buizert2007,Eduardo2012,Franke2011,Kim2013,Chang2013}
In the present work, we find that the Rashba SOC introduces a novel Kondo screening channel of the magnetic impurity in a superconductor. The basic idea is summarized as follows.
In an s-wave superconductor, the Rashba SOC creates a spin-triplet pairing state and suppresses the superconductivity by mixing the spin and orbit angular momentum \cite{Sigrist1991,Gorkov2001,Frigeri2004}. When the Rashba interaction $\varepsilon_{R}$ is large enough, the electrons in Cooper pairs would undergo a total angular momentum conserved quantum state transition (QST) process, such as the transition $\left\vert 0,\uparrow \right\rangle \rightarrow \left\vert 1,\downarrow\right\rangle $ as shown in Fig.\ref{fig1}\,(b), where $\left\vert m,s\right\rangle$ denotes a state with the orbital quantum number $l_z=m$ and the spin $s_z=s(\uparrow, \downarrow)$. Then, the spin-down electron ($\left\vert 1,\downarrow\right\rangle$) replaces the spin-up electron on impurity level $\varepsilon_s$ by the double or empty occupied virtual states, see Fig.\ref{fig1}\,(c). Accordingly, the spin-up electron can tunnel into the superconductor and neutralizes the hole ($\left\vert 0,\uparrow\right\rangle$). As a result, the local moment of the magnetic impurity is effectively flipped as shown in Fig.\ref{fig1}\,(d). In the same way, the spin-down electron on impurity level can be replaced by a spin-up electron excited from the superconductor by this process ($\left\vert 0,\downarrow \right\rangle \rightarrow\left\vert -1,\uparrow\right\rangle$). The coherent superposition of these processes provides a novel Kondo screening channel to the magnetic impurity in a superconductor,
which is confirmed by the Rashba promoted QPT between the magnetic doublet and Kondo singlet ground states.

The paper is organized as follows. In Sec. \ref{model}, we present the model and formalism as our starting point for explaining our picture. In Sec. \ref{result} we show the results obtained by solving the Green's functions in a self-consistent way. Some discussion has also been presented. Finally, a brief conclusion is devoted to Sec. \ref{conclusion}.

\section{The model and formalism} \label{model}
The Hamiltonian of the system reads
\begin{eqnarray}
H=H_{0}+H_{d}+H_{V},  \label{H}
\end{eqnarray}
where $H_{0}=\sum\limits_{\mathbf{k,}s}\varepsilon( \mathbf{k}) c_{%
\mathbf{k,}s}^{\dagger }c_{\mathbf{k,}s}-\Delta\sum\limits_{\mathbf{k}}( c_{\mathbf{k,}\downarrow }^{\dagger }c_{-\mathbf{k,}\uparrow}^{\dagger }+h.c.)+\lambda _{R}\sum\limits_{\mathbf{k}}k( e^{-i\theta _{k}}c_{\mathbf{k,%
}\downarrow }^{\dagger }c_{\mathbf{k,}\uparrow }+h.c.)$
describes the electrons in the conduction band. $\varepsilon (
\mathbf{k}) =\frac{\hbar ^{2}\mathbf{k}^{2}}{2m^{\star}}-\mu$ is the
dispersion of conduction electrons,$m^{\star}$ is the effective electron mass, $\mu $ is the gate-tunable chemical
potential, $\Delta$ is the proximity induced energy gap, $\lambda _{R}$ is the
Rashba SOC strength parameter, and $k=|\mathbf{k}|$, $\mathbf{k}=k_{x}\hat{\mathbf{x}}%
+k_{y}\hat{\mathbf{y}}$, $\theta _{k}=\text{atan}(k_{x}/k_{y})$ is the polar
components of the $\mathbf{k}$ \cite{Zitko2011,Wong2016}. The operator $c_{\mathbf{\pm k,%
}s}^{\dagger }$ ($c_{\mathbf{\pm k,}s}$) stands for the creation (annihilation)
of an electron with the momentum $\mathbf{\pm k}$
and the spin $s$ $(=\uparrow ,\downarrow )$.
The magnetic impurity can be described by the Hamiltonian
$H_{d}=\sum_{s}\varepsilon _{s}d_{s}^{\dagger }d_{s}+Un_{\uparrow}n_{\downarrow }$
where $\varepsilon_{s}$ is the impurity level, $U$ is the Coulomb repulsion, $%
d_{s}^{\dagger }$ ($d_{s}$) is the creation (annihilation) operator of $d$%
-electrons, and $n_{s}=d_{s}^{\dagger }d_{s}$. The hybridization between the
impurity and the superconductor is $H_{V}=\sum_{\mathbf{k,}s}( V_{\mathbf{k,}s}c_{\mathbf{k,}s}^{\dagger
}d_{s}+h.c.)$, $V_{\mathbf{k}s}$ is the hybridization amplitude, and we assume $V_{\mathbf{k}s}=V$ is a constant for simplification.
The summation of $\mathbf{k}$ can be transformed into the integration in polar coordinates, $\sum_{%
\mathbf{k}}=\frac{S}{(2\pi) ^{2}}\int kdkd\theta _{k}$, and $%
S\equiv 1$.

In the presence of the SOC, it is convenient to introduce the angular momentum basis for the conduction band
electrons $c_{\mathbf{k,}s}=\sqrt{\frac{2\pi }{k}}\sum_{m=-\infty }^{\infty
}e^{im\theta _{k}}c_{k,s}^{m}$, with $m$ the orbital angular momentum quantum number \cite{Zarea2012}. The canonical
anti-commutation relation of the conduction electron under the angular momentum basis is $[c_{k,s}^{m},(c_{k^{\prime },s^{\prime
}}^{m^{\prime }})^{\dagger }]_{+}=\delta (k^{\prime }-k)\delta _{s^{\prime
}s}\delta _{m^{\prime }m}$\cite{Zitko2011,Wong2016}. In order to diagonalize the Rashba interaction
term, a canonical transformation of the fermionic
operators $\tilde{c}_{h,k,m+1/2}=\frac{1}{\sqrt{2}}( c_{k,\uparrow
}^{m}+hc_{k,\downarrow }^{m+1})$ has been introduced \cite{Zarea2012}, 
and here $h=\pm 1$ is the helicity quantum number.
The Hamiltonian Eq.(\ref{H}) can be rewritten as
\begin{eqnarray}
H &=&\sum_{h,j}\int dk\varepsilon _{h,k}\tilde{c}_{h,k,j}^{\dagger }\tilde{c}%
_{h,k,j}+\sum_{j}\varepsilon _{j}d_{j}^{\dagger }d_{j}+Un_{\uparrow
}n_{\downarrow }  \notag \\
&&-\frac{\Delta}{2}\sum\limits_{h,h^{\prime },j}\int dk(h\tilde{c}%
_{h,k,j}^{\dagger }\tilde{c}_{h^{\prime },-k,-j}^{\dagger }+h.c.)  \notag \\
&&+\sum_{h,j}\int dk\tilde{V}_{k}h^{-(j-1/2)}(\tilde{c}_{h,k,j}^{\dagger
}d_{j}+h.c.),  \label{Hnew}
\end{eqnarray}
where $\tilde{V}_{k}=\sqrt{k/(4\pi )}V$, $\varepsilon_{h,k}=\varepsilon(k)+h\lambda _{R}k$,
and the total angular momentum $j=j_z=m+s_{z}$. Here, we denote the operator $d_{s}\rightarrow d_{j}$ of the impurity level.
The Rashba interaction leads to an indirect exchange coupling between magnetic impurity with spin $s = \pm 1/2$ and conduction
electrons with different spin and orbital states \cite{Malecki2007,Wong2016},
e.g. the magnetic spin $d_{j}$ couples the conduction electrons
$c_{k,\uparrow}^{m}$ and $c_{k,\downarrow }^{m+1}$ in Eq.(\ref{Hnew}).
In this case, the magnetic impurity with spin $\pm\frac{1}2$ couples the conduction
electrons with the orbital quantum number $m=0, \pm1$ due to the fact that $j_{z}$ is a conserved quantity. 

The Hamiltonian in Eq.(\ref{Hnew}) can be systematically treated by the equation of motion
approach, which gives qualitative descriptions for the Kondo effect \cite{Lacroix1981,Luo1999}
and the interplay between Kondo effect and superconductivity \cite{Li2014,Li2016}.
In frequency space, the Nambu Green's function (GF) can be expressed with Dyson equation
\begin{equation}
\hat{G}_{d,j}(\omega) =\hat{G}_{d,j}^{0}(\omega)+
U\hat{F}_{d,j}\left( \omega \right)\hat{G}_{d,j}^{0}\left(
\omega \right), \label{GF}
\end{equation}
where the noninteracting GF reads
$\hat{G}_{d,j}^{0}(\omega) =(\hat{I}\omega -\sigma _{z}\text{diag}%
(\varepsilon _{j},\varepsilon _{-j})-\hat{\Sigma}^{0}( \omega ))^{-1}$.
The elements of noninteracting self-energy are $\hat{\Sigma}_{11(22)}^{0}( \omega
)=\frac{1}{\pi }\sum_{h}\int \Gamma _{h}( \varepsilon _{h,k})
( \omega+\varepsilon_{h,k})\beta _{h,k}(\omega)d\varepsilon _{h,k}$ and
$\hat{\Sigma}_{12(21)}^{0}( \omega) =-\frac{\Delta}{\pi }$sign($j$)$
\sum_{h}\int \Gamma _{h}(\varepsilon _{h,k}) \beta
_{h,k}(\omega)d\varepsilon _{h,k}$,
where the notations $\beta _{h,k}(\omega) =
\frac{g_{h,k}(\omega) }{( \omega -\varepsilon
_{h,k})(\omega +\varepsilon _{h,k}) g_{h,k}(\omega) -\Delta^{2}}$
and $g_{h,k}(\omega) =\frac{\omega ^{2}-\varepsilon _{-h,k}^{2}}{\omega
^{2}-\varepsilon _{k}^{2}}$ (see appendix). The coupling is $\Gamma _{h}\left( \varepsilon _{h,k}\right)=
\Gamma_0\rho _{h}(\varepsilon _{h,k})$ with $\Gamma_0=\pi \left\vert V\right\vert ^{2}/2D$,
and the density of states
\begin{small}
\begin{equation}
\rho _{h}\left( \varepsilon _{h,k} \right) =\left\{
\begin{array}{c}
\frac{m^{\star}}{2\pi \hbar ^{2}}\frac{\varepsilon _{R}}{\sqrt{\varepsilon
_{R}\left(\varepsilon_{h,k} -\varepsilon _{0}\right) }}\delta _{h,-1;}
E_{0}-\varepsilon _{R}<\varepsilon_{h,k}<E_{0} \\
\frac{m^{\star}}{2\pi \hbar ^{2}}\left( \frac{1}{2}-\frac{h\varepsilon _{R}}{2\sqrt{
\varepsilon _{R}\left(\varepsilon_{h,k}-\varepsilon _{0}\right) }}\right)
;E_{0}<\varepsilon_{h,k} <D_{h} \\
0\text{ \ \ \ \ \ \ \ \ \ \ \ \ \ \ \ \ \ \ \ \ \ \ otherwise},
\end{array}
\right.   \label{Rho}
\end{equation}
\end{small}
where $D_{h}=D+2h\sqrt{\left( D-E_{0}\right) \varepsilon _{R}}$ is the
helicity dependent band-width, $D$ and $E_{0}$ are the half band-width and
the bottom of conduction band without spin-orbit coupling,
respectively, $\varepsilon _{R}=\frac{k_{0}^{2}}{2m^{\star}}$ is the Rashba energy with
$k_{0}=\frac{m^{\star}\lambda _{R}}{\hbar }$\cite{Zitko2011,Wong2016}.

In Eq.(\ref{GF}), the matrix $\hat{F}_{d,j}\left( \omega \right)$ can be read explicitly
\begin{equation}
\hat{F}_{d,j}\left( \omega \right) =\left(
\begin{array}{cc}
\langle \langle d_{j}n_{-j};d_{j}^{\dagger }\rangle \rangle & \langle
\langle d_{j}n_{-j};d_{-j}\rangle \rangle \\
-\langle \langle d_{-j}^{\dagger }n_{j};d_{j}^{\dagger }\rangle \rangle &
-\langle \langle d_{-j}^{\dagger }n_{j};d_{-j}\rangle \rangle
\end{array}\right). \label{Fmatrix}
\end{equation}
However, it is difficult to exactly calculate the elements in above matrix.
In order to qualitatively obtain the Kondo physics, we take the Lacroix's
scheme to treat the diagonal elements in $\hat{F}_{d,j}\left(\omega\right)$ \cite{Lacroix1981,Li2016}.
While the off-diagonal elements are approximately given by Hartree-Fock approximation ($U\gg\Delta$), see detail in appendix.
After some straightforward calculations, we obtain the GF
\begin{equation}
\lbrack \hat{G}_{d,j}(\omega) ]_{11}=\frac{1+UO_{j}(
\omega) +( 1+UP_{j}( \omega ))\Pi_{21}( \omega )}{%
\omega -\varepsilon _{j}-\hat{\Sigma}_{11}^{0}( \omega)
-UQ_{j}(\omega) },  \label{Gd11}
\end{equation}
where the notations $O_{j}(\omega)$, $P_{j}(\omega)$, $Q_{j}(\omega)$ are shown explicitly in appendix, see Eq.(\ref{Oj})-(\ref{Qj}).
And the notation $\Pi_{21}( \omega)= \hat{\Sigma}_{21}^{0}( \omega) [\hat{G}_{d,j}( \omega) ]_{21}$.
The anomalous GF obtained is
\begin{equation}
\lbrack \hat{G}_{d,j}( \omega ) ]_{21}=\frac{\hat{\Sigma}%
_{21}^{0}( \omega ) +U\langle d_{j}^{\dagger }d_{-j}^{\dagger
}\rangle }{\omega +\varepsilon _{-j}+U\langle n_{j}\rangle -\hat{%
\Sigma}_{11}^{0}( \omega )}[\hat{G}_{d,j}( \omega) ]_{11}, \label{G21}
\end{equation}
where the occupation $\langle n_{j}\rangle =-\frac{1}{\pi }\int
f( \omega) $Im$[\hat{G}_{d,j}( \omega ) ]_{11}d\omega $%
, $f( \omega) $ is the Fermi distribution function.
The pairing correlation function $\langle d_{j}^{\dagger
}d_{-j}^{\dagger }\rangle $ can be evaluated by $\langle d_{j}^{\dagger
}d_{-j}^{\dagger }\rangle =-\frac{1}{\pi }\int f( \omega) $Im$[%
\hat{G}_{d,j}( \omega) ]_{21}d\omega.$
Then, the GFs $[\hat{G}_{d,j}( \omega) ]_{11}$ and $[\hat{G}_{d,j}( \omega ) ]_{21}$
can be calculated self-consistently with above formulism.
The phenomena introduced by magnetic moment in superconductor
can be qualitatively discussed based on the numerical results.

\section{The numerical results and discussions}\label{result}

\begin{figure}[tbp]
\includegraphics[clip=true,width=1.0\columnwidth]{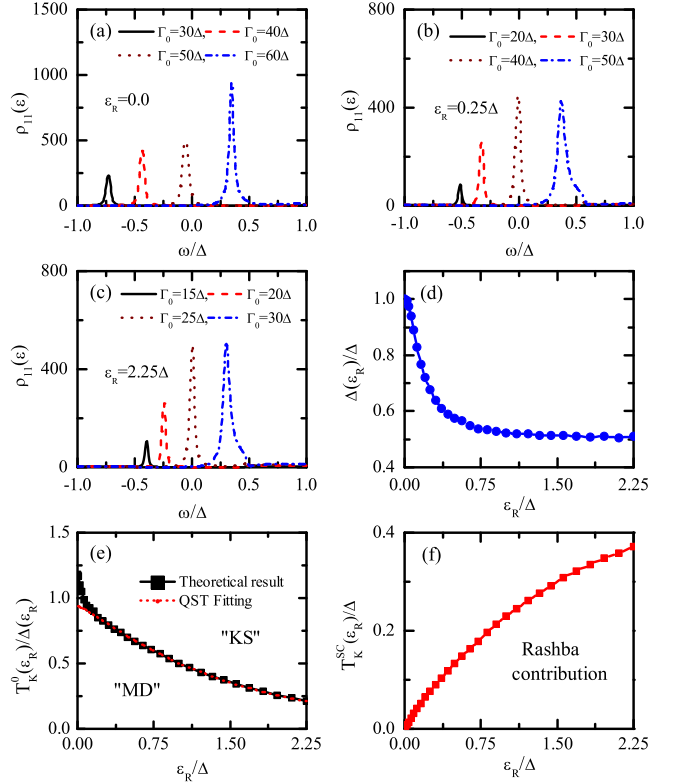}
\caption{(a)-(c) The YSR bound states ($E_{YSR}$) are tuned from below to above the Fermi level by increasing the tunneling amplitude $\Gamma_0$ with different Rashba SOC energy $\varepsilon_R=0.0,0.25\Delta,2.25\Delta$, respectively. The quantum phase transition (QPT) between the magnetic doublet (MD) and Kondo singlet (KS) ground states occurs when the YSR states get across the Fermi level. (d) The Rashba SOC suppressed superconducting energy gap as a function of $\varepsilon_{R}$. (e) The phase boundary between the MD and KS phases is scaled by $T^{0}_K(\varepsilon_{R})/\Delta\left(\varepsilon_{R}\right)$ see the square solid line.
The suppression is attributed to the Rashba SOC induced Kondo screening process,
which is confirmed by a quantum states transition (QST) model as shown the red dash-dotted line.
(f) The Kondo temperature $T^{SC}_K(\varepsilon_R)$ characterizing the Rashba induced Kondo screening process in superconductor.
Other parameters used are the impurity level $\varepsilon_{j}=-10\Delta$, the Coulomb interaction $U=100\Delta$, the half-band width $D=25\Delta$, and the temperature $T=0$.}
\label{fig2}
\end{figure}

The YSR bound state, in local density of states (DOS) $\rho_{11}(\omega)=-\frac{1}{\pi }\sum_{j}$Im$[\hat{G}_{d,j}(\omega)]_{11}$,
reflects directly the interplay between the Kondo screening and superconductivity \cite{Zhu2006,Kim2015,Ruby2016,Jellinggaard2016}.
The YSR bound state can be tuned from below ($E_{YSR}<0$) to above ($E_{YSR}>0$) the Fermi level by increasing the tunneling amplitude $\Gamma_0$ [see Figs.\ref{fig2}\,(a)-(c)].
The QPT between the magnetic doublet state and the Kondo screened singlet state takes place for the YSR bound state crossing the Fermi level ($E_{YSR}=E_F=0$).
The characteristic energy scale of the phase transition is $T^{0}_K/\Delta=c$, where $c(\sim 1)$ is a constant \cite{Siano2004,Li2014,Clerk2000,Rozhkov2000,Choi2004}. The normal state Kondo temperature can be obtained from Haldane's scaling theory on asymmetric Anderson model\cite{Haldane1978}, $T^{0}_K\approx\Gamma\text{exp}(\pi\varepsilon_j/2\Gamma)$, $\Gamma$ is the coupling amplitude. In the absence of Rashba interaction, in Fig.\ref{fig2}\,(a), the quantum phase transition occurs around $T^{0}_K/\Delta=1.17$. This result is in agreement with the experimental observations\cite{Island2017}.
In the presence of Rashba SOC, in Figs.\ref{fig2}\,(b) and \ref{fig2}(c), the QPT between the ground states takes place at $T^{0}_K(\varepsilon_{R})/\Delta(\varepsilon_{R})=0.83$ and $ 0.21$ for  $\varepsilon_R=0.25\Delta$ and $2.25\Delta$, respectively.
Where the normal state Kondo temperature is also given by Haldane's formula $T^{0}_K(\varepsilon_{R})\approx\Gamma(\varepsilon_{R})\text{exp}(\pi\varepsilon_j/2\Gamma(\varepsilon_{R}))$, the coupling amplitude is approximately given by $\Gamma(\varepsilon_{R})=\Gamma_{0}\rho(E_F)$, because the density of states $\rho(\varepsilon)=\sum_{h}\rho_h(\varepsilon)$ around the Fermi level is a constant when the Rashba energy $\varepsilon_R$ is not large enough [see Eq.(\ref{Rho})].
And the superconducting energy gap $\Delta(\varepsilon_{R})$ is suppressed due to the mixing of spin and orbit angular momentum of conduction electrons [see Fig.\ref{fig2}\,(d)]. In Fig.\ref{fig2}\,(e), we plot the phase diagram dominated by the competition between Kondo effect and superconductivity. The phase boundary between the magnetic doublet (MD) state and Kondo singlet (KS) state is characterized by the energy scale $T^{0}_K(\varepsilon_{R})/\Delta(\varepsilon_{R})$. Instead of a constant without Rashba SOC, the energy scale $T^{0}_K(\varepsilon_{R})/\Delta(\varepsilon_{R})$ is shown suppressed with the increase of $\varepsilon_{R}$.
This fact is attributed to the Rashba induced Kondo screening channel of magnetic impurity [see Figs.\ref{fig1}\,(b)-(d)].

In the following, we theoretically analyze the Rashba induced Kondo screening process.
Here, we denote that the Kondo screening is essentially determined by the prerequisite QST between different spin and orbital states [see Fig.\ref{fig1}\,(b)].
Then, the Kondo temperature can be obtained from the intensity of QST processes.
In the presence of Rashba SOC, the spin and the orbital angular momentum are not conserved quantities.
Then, the conduction electrons undergo some total angular momentum preserved QST processes, such as
$\vert 0,\uparrow\rangle \leftrightarrow\vert 1,\downarrow\rangle $ and $\vert 0,\downarrow
\rangle \leftrightarrow \vert -1,\uparrow \rangle $ as seen in Fig.\ref{fig1}\,(b).
The probability of these processes can be approximately calculated by the perturbation theory, $\lambda _{R}k$ is the perturbation.
Therefore, we neglect the superconducting term, and rewrite the Hamiltonian of conduction electrons in the angular momentum basis, $H_{0}=\tilde{H}_{0}+\tilde{H}_{0}^{\prime }$ with $\tilde{H}_{0}=\sum_{m,s}\int dk\varepsilon_{k}c_{k,s}^{m\dagger}c_{k,s}^{m}$ and the perturbation term
$\tilde{H}_{0}^{\prime }=\int dk\lambda _{R}k( c_{k\uparrow }^{-1\dagger
}c_{k\downarrow }^{0}+c_{k\downarrow }^{0\dagger }c_{k\uparrow
}^{-1}+c_{k\uparrow }^{0\dagger }c_{k\downarrow }^{1}+c_{k\downarrow
}^{1\dagger }c_{k\uparrow }^{0})$, ($j=\pm 1/2$).
The transition probability of QST processes can be easily obtained with $P\propto\lambda _{R}^{2}k^{2}$.
On one hand, the QST processes would suppress the spin-singlet pairing ground state
$\vert \Psi_{G}\rangle =\Pi _{k}(u_{k}+\upsilon _{1k}c_{k\uparrow }^{0\dagger}c_{-k\downarrow }^{0\dagger }
+\upsilon _{2k}c_{k\uparrow }^{-1\dagger}c_{-k\downarrow }^{1\dagger })\vert \phi _{0}\rangle $;
on the other hand, it could even excite an electron from Cooper pairs at large Rashba interaction case.
Then, the electron directly contributes to the Kondo screening of local spin [see Figs.\ref{fig1}\,(b)-(d)].
Taking account of this effect, the effective coupling becomes $\Gamma(\varepsilon_{R}) \rightarrow \Gamma(\varepsilon_{R}) +\delta \Gamma $, and $\delta \Gamma \propto P$.
Then, the Kondo temperature increases due to the QST processes. One can easily obtain the ground states QPT taking place at
$T_{K}^{0}(\varepsilon_{R})/\Delta ( \varepsilon_{R}) =c/(1+x)$, with $x=\frac{2\alpha \varepsilon_{R}}{\Gamma(\varepsilon_{R}) }[1+\frac{\pi\vert\varepsilon_{d}\vert }{2\Gamma(\varepsilon_{R}) }(1+\frac{2\alpha\varepsilon_{R}}{\Gamma(\varepsilon_{R})})]$.
By fitting the numerical results marked as red dotted line in Fig.\ref{fig2}\,(e), the parameters obtained are $c=0.94$ and $\alpha =0.5$.
The Kondo temperature characterizing the Rashba induced Kondo screening process in superconductor is $T_{K}^{SC}(\varepsilon_{R})=\frac{c\Delta(\varepsilon_{R}) x}{1+x}$,
which shows a significant enhancement with the increase of $\varepsilon_{R}$ due to the increasing of QST processes [see Fig.\ref{fig2}\,(f)].

\begin{figure}[tbp]
\includegraphics[clip=true,width=1.0\columnwidth]{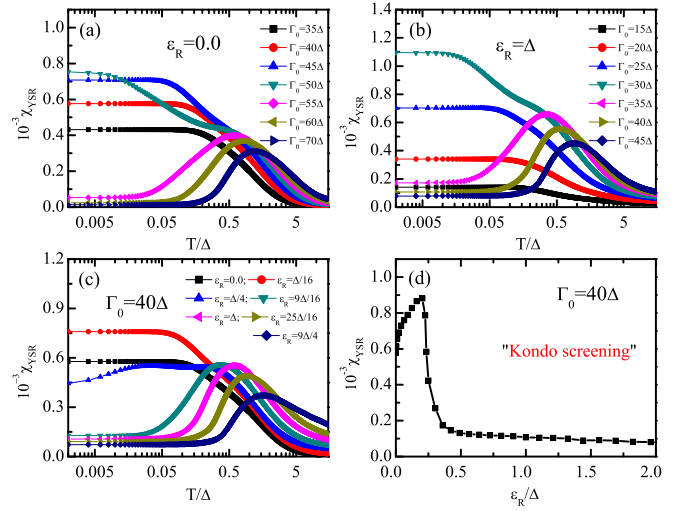}
\caption{(a)-(b) The temperature dependent susceptibility of YSR state varying with the tunneling amplitude $\Gamma_0$ with different spin-orbit coupling strengths $\varepsilon_R=0.0, \Delta$,respectively. (c) Temperature dependent susceptibility of YSR state suppressed by the increasing $\varepsilon_R$, and $\Gamma_0=40\Delta$. (d) Enhancement and suppression of susceptibility by increasing the Rashba interaction at $T=0$ and $\Gamma_0=40\Delta$. The other parameters used are the same with those in Fig.\ref{fig2}.}
\label{fig3}
\end{figure}

The temperature dependence of susceptibility reflects directly the Kondo screening behavior of magnetic impurity.
In a superconductor, the Kondo screening of magnetic moment can be discussed by the susceptibility of YSR bound state,
because it possesses effective moment\cite{Li2016,Domanski2016}.
Therefore, the Rashba induced Kondo screening can be discussed by the susceptibility of YSR bound states.
Here, we define the susceptibility of YSR bound states
$\chi_{_{YSR}}=\frac{g\mu_B(n_{Aj}-n_{A-j})}{\mathcal{H}}\vert_{\mathcal{H}\rightarrow 0}$,
where the occupation is $n_{Aj}=-\frac{1}{\pi}\int^{+\Delta}_{-\Delta} f(\omega) \text{Im}[\hat{G}_{d,j}(\omega)]_{11}d\omega$,
$\mu_B$ is the Bohr magneton, $g$ is the Land\'e factor, $\mathcal{H}$ is a weak magnetic field, and we take $\mu_B=g=1$.
In Fig.\ref{fig3} (a), in the absence of Rashba interaction the susceptibility is enhanced in magnetic doublet ground state by increasing the coupling $\Gamma_0$ ($\Gamma_0=35\Delta,40\Delta,45\Delta,50\Delta$) at low temperatures,
which reflects the development of the effective moment in YSR bound states.
Once the system enters into the Kondo singlet regime, the moment is significantly quenched ($\Gamma_0\geq 55\Delta$),
and the susceptibility shows typical temperature-dependent behavior of Kondo screening.
At high temperatures, the susceptibility satisfies the Curie's law $\chi_{_{YSR}}\propto \frac{1}T$.
Similar tendency is observed for $\varepsilon_{R}=\Delta$ [see Fig.\ref{fig3}\,(b)], where the critical point is promoted to $\Gamma_0\approx 35\Delta$.
In Fig.\ref{fig3}\,(c), we show the susceptibility of YSR bound state is significantly suppressed by increasing the Rashba SOC from $\varepsilon_{R}=0.0$ to $\varepsilon_{R}=9\Delta/4$.
It indicates that the Rashba interaction could contribute to the Kondo screening of magnetic impurity in superconductor.
In Fig.\ref{fig3}\,(d), the susceptibility is enhanced by the Rashba interaction for $\varepsilon_{R}<0.25\Delta$ at zero temperature, while it is rapidly suppressed by the Rashba interaction for $\varepsilon_{R}\geq 0.25\Delta$.
In former case, the enhancement of susceptibility is attributed to the suppression of superconductivity, and the local moment is shifted from the impurity to YSR bound states.
The ground state is a magnetic doublet state when the energy scale $(T^{0}_{K}(\varepsilon_{R})+T^{R}_{K}(\varepsilon_{R}))/\Delta({\varepsilon_{R}})< 1$. In large $\varepsilon_R$ cases, the Rashba induced QST processes directly contribute to the Kondo screening of the magnetic impurity. Thus, the effective moment of YSR bound states is rapidly quenched by the Kondo screening. Then, the ground state of the system is a Kondo singlet state for $(T^{0}_{K}(\varepsilon_{R})+T^{R}_{K}(\varepsilon_{R}))/\Delta({\varepsilon_{R}})>1$. At the critical point $\varepsilon_{R}\approx0.25\Delta$, the Kondo temperature and the superconducting energy gap satisfies the relationship $(T^{0}_{K}(\varepsilon_{R})+T^{R}_{K}(\varepsilon_{R}))/\Delta({\varepsilon_{R}})\approx1$,
which suggests that the Rashba SOC directly contributes to the Kondo screening of the magnetic impurity.

The Rashba enhanced Kondo screening of magnetic impurity in normal metal has been predicted in many works\cite{Zarea2012,Isaev2012,Isaev2015,Wong2016,Sousa2016,Chen2016}.
For example, the Dzyaloshinsky-Moriya (DM) interaction introduced by Rashba SOC leads to an exponential enhancement of the Kondo temperature, which originates from the exchange coupling between magnetic impurity and conduction electrons with different orbit angular momentum.\cite{Zarea2012} Similarly, the Kondo screening originating from purely orbital states was predicted for a magnetic impurity on the surface of topological insulator with strong spin-orbit coupling\cite{Isaev2015}. Different to these papers, in the present work we found that the Rashba interaction not only suppresses the superconducting energy gap, but also introduces an novel Kondo screening channel. The Kondo screening involves the exchange coupling between the local spin and conduction electrons with different spin and orbital states [see Fig.\ref{fig1}\,(c)]. In principle, the exchange coupling between the magnetic impurity and conduction electrons with different spin and orbital states can be introduced by atomic spin-orbit coupling. Our work may provide some insights into the Majorana bound states at the end of magnetic atoms chain on superconductor with strong spin-orbit coupling.

\section{Conclusion}\label{conclusion}
In the present paper we find that the Rashba SOC generates an additional channel to screen the magnetic impurity in superconductor. The Kondo screening originates from the exchange coupling between magnetic impurity and conduction electrons with different spin and orbital states. Consequently, the energy scale $T^{0}_K(\varepsilon_{R})/\Delta(\varepsilon_R)$, characterizing the phase transition between the magnetic doublet and Kondo singlet ground states, decreases with the increasing $\varepsilon_R$. This result sheds a novel insight on the interplay between the Kondo effect and superconductivity. This result can be observed by the scanning tunneling microscopy and the magnetic susceptibility measurements. Our work may be useful in understanding the physics emerged in noncentrosymmetric superconductors.

\section{Acknowledgement}
Lin Li acknowledges useful discussion with Dr. Hua Chen and Dr. Wei Chen. This work is supported by the National Key Research and Development Program of China (No.2016YFA0300300), and NSFC (No.11604138, No.11674151, No.11547110, No.11325417, No.11674139), PCSIRT (No.IRT-16R35) of China, and Guangdong Natural Science Foundation 2014A030310137.

\section{Appendix}
In this appendix, we present the main steps to treat the Hamiltonian in Eq.(%
\ref{Hnew}) by the equation of motion (EOM) approach. The retarded GF can be read
\cite{Zubarev1960}
{\small
\begin{equation}
\omega \langle \langle A;B\rangle \rangle =\langle \lbrack A,B]_{+}\rangle
+\langle \langle \lbrack A,H]_{-};B\rangle \rangle ,  \label{EM}
\end{equation}
}
where the subscript $\pm $ stands for the anti-commutation (commutation)
relationship, and {\small $\langle \langle A;B\rangle \rangle$} denotes the
retarded GF composed by the operators $A$ and $B$.

By substituting Eq.(\ref{Hnew}) into Eq.(\ref{EM}), we obtain the Green functions
composed by the creation (annihilation) operators of local and conduction electrons
{\small
\begin{eqnarray}
\omega \left\langle \left\langle d_{j};B\right\rangle \right\rangle
&=&\langle \left[ d_{j};B\right] _{+}\rangle +\varepsilon _{j}\left\langle
\left\langle d_{j};B\right\rangle \right\rangle +U\left\langle \left\langle
d_{j}n_{-j};B\right\rangle \right\rangle  \notag \\
&&+\sum_{h}\int dkh^{-\left( j-1/2\right) }\tilde{V}_{k}\langle \langle
\tilde{c}_{h,k,j};B\rangle \rangle ,  \label{Gdb}
\end{eqnarray}
}
{\small
\begin{eqnarray}
\omega \left\langle \left\langle \tilde{c}_{h,k,j};B\right\rangle
\right\rangle &=&\langle \left[ \tilde{c}_{h,k,j};B\right] _{+}\rangle
+\varepsilon _{h,k}\left\langle \left\langle \tilde{c}_{h,k,j};B\right%
\rangle \right\rangle  \notag \\
&&+h^{-\left( j-1/2\right) }\tilde{V}_{k}\left\langle \left\langle
d_{j};B\right\rangle \right\rangle  \notag \\
&&+\frac{\Delta }{2}\delta _{j,-1/2}\sum\limits_{h^{\prime }}h^{\prime
}\langle \langle \tilde{c}_{h^{\prime },-k,-j}^{\dagger };B\rangle \rangle  \notag \\
&&-\frac{\Delta }{2}\delta _{j,1/2}h\sum\limits_{h^{\prime }}\langle \langle
\tilde{c}_{h^{\prime },-k,-j}^{\dagger };B\rangle \rangle ,
\label{GccB}
\end{eqnarray}
}
{\small
\begin{eqnarray}
\omega \langle \langle \tilde{c}_{h^{\prime },-k,-j}^{\dagger };B\rangle
\rangle &=&\langle \lbrack \tilde{c}_{h^{\prime },-k,-j}^{\dagger
};B]_{+}\rangle -\varepsilon _{h^{\prime },-k}\langle \langle
\tilde{c}_{h^{\prime },-k,-j}^{\dagger };B\rangle \rangle
 \notag \\
&&-h^{\prime \left( j+1/2\right) }\tilde{V}_{k}\langle \langle
d_{-j}^{\dagger };B\rangle \rangle  \notag \\
&&-\frac{\Delta }{2}\delta _{j,1/2}\sum\limits_{h}h\left\langle \left\langle
\tilde{c}_{h,k,j};B\right\rangle \right\rangle  \notag \\
&&+\frac{\Delta }{2}\delta _{j,-1/2}h^{\prime }\sum\limits_{h}\left\langle
\left\langle \tilde{c}_{h,k,j};B\right\rangle \right\rangle ,  \label{Gc+B}
\end{eqnarray}
}
and
{\small
\begin{eqnarray}
\omega \langle \langle d_{-j}^{\dagger };B\rangle \rangle
&=&\langle \lbrack
d_{-j}^{\dagger };B]_{+}\rangle -\varepsilon _{-j}\langle \langle
d_{-j}^{\dagger };B\rangle \rangle -U\langle \langle d_{-j}^{\dagger
}n_{j};B\rangle \rangle  \notag \\
&&-\sum_{h}\int dkh^{(j+1/2)}\tilde{V}_{k}\langle \langle \tilde{c}%
_{h,k,-j}^{\dagger };B\rangle \rangle .  \label{Gd+B}
\end{eqnarray}
}
After some straightforward algebraic calculations, we obtain the GFs:
{\small
\begin{eqnarray}
\omega \langle \langle d_{j};B\rangle \rangle
&=&\langle \lbrack d_{j};B]_{+}\rangle +\varepsilon _{j}\langle \langle
d_{j};B\rangle \rangle +U\langle \langle d_{j}n_{-j};B\rangle \rangle  \notag \\
&&+\text{sign}(j)\frac{\Delta }{2}\int dk\frac{|\tilde{V}_{k}|^{2}\tilde{g}%
\left( k,\omega \right) }{1-\frac{\Delta ^{2}}{4}g\left( k,\omega \right) }%
\langle \langle d_{-j}^{\dagger };B\rangle \rangle  \notag \\
&&+\sum\limits_{h}\int dk\frac{\tilde{V}_{k}|^{2}\langle \langle d_{j};B\rangle \rangle}{\left( \omega
-\varepsilon _{h,k}\right) \left( 1-\frac{\Delta ^{2}}{4}g(k,\omega) \right) }
,  \label{DB1}
\end{eqnarray}
}
and
{\small
\begin{eqnarray}
\omega \langle \langle d_{-j}^{\dagger };B\rangle \rangle
&=&\langle \lbrack d_{-j}^{\dagger };B]_{+}\rangle -\varepsilon _{-j}\langle
\langle d_{-j}^{\dagger };B\rangle \rangle -U\langle \langle d_{-j}^{\dagger
}n_{j};B\rangle \rangle  \notag \\
&&+\text{sign}(j)\frac{\Delta }{2}\int dk\frac{|\tilde{V}_{k}|^{2}\tilde{g}%
\left( k,\omega \right) }{1-\frac{\Delta ^{2}}{4}g\left( k,\omega \right) }%
\langle \langle d_{j};B\rangle \rangle
\notag \\
&&+\sum\limits_{h}\int dk\frac{|\tilde{V}_{k}|^{2}\langle \langle d_{-j}^{\dagger };B\rangle \rangle}{\left( \omega
+\varepsilon _{h,-k}\right) \left( 1-\frac{\Delta ^{2}}{4}g(k,\omega) \right) } ,  \label{DB2}
\end{eqnarray}
}
where the notations {\small$g\left( k,\omega \right) =\sum\limits_{h,h^{\prime}}\frac{1}{%
\omega -\varepsilon _{h,k}} \frac{1}{\omega
+\varepsilon _{h^{\prime },-k}}$} and {\small$\tilde{g}\left( k,\omega \right)
=\sum\limits_{h,h^{\prime}}\frac{h}{\omega -\varepsilon _{h,k}}
\frac{h^{\prime }}{\omega +\varepsilon _{h^{\prime },-k}}.$} Here, we neglect
the terms in the order {\small$\Delta ^{2}$}.

By assuming {\small$B=(d_{-j},d_{j}^{\dagger })$} in Eqs.(\ref{DB1}) and (\ref{DB2}),
one can obtain four equations of the Nambu GF. Then, we rewrite Nambu GF in matrix presentation
{\small
\begin{equation}
\hat{G}_{d,j}\left( \omega \right) =\hat{G}_{d,j}^{0}\left( \omega \right) + U\hat{F}_{d,j}(\omega )\hat{G}_{d,j}^{0}(\omega ) ,  \label{GmatrixApp}
\end{equation}
}
where
{\small
\begin{equation}
\hat{G}_{d,j}^{0}\left( \omega \right) =\left( \hat{I}\omega -\sigma _{z}%
\text{diag}(\varepsilon _{j},\varepsilon _{-j})-\hat{\Sigma}^{0}\left(
\omega \right) \right) ^{-1}  \label{G0matrixAPP}
\end{equation}
}
is the noninteracting GF, {\small $\hat{\Sigma}_{j}^{0}\left( \omega \right) $} is
the noninteracting self-energy, and the notation
{\small
\begin{equation}
\hat{F}_{d,j}\left( \omega \right) =\left(
\begin{array}{cc}
\langle \langle d_{j}n_{-j};d_{j}^{\dagger }\rangle \rangle & \langle
\langle d_{j}n_{-j};d_{-j}\rangle \rangle \\
-\langle \langle d_{-j}^{\dagger }n_{j};d_{j}^{\dagger }\rangle \rangle &
-\langle \langle d_{-j}^{\dagger }n_{j};d_{-j}\rangle \rangle%
\end{array}%
\right)  \label{Fmatrix}
\end{equation}
}
involves some higher order GFs.

The diagonal elements of the noninteracting self-energy can be transformed into
{\small
\begin{equation}
\hat{\Sigma}_{11(22)}^{0}\left( \omega \right) =\frac{1}{\pi }\sum_{h}\int
\Gamma _{h}\left( \varepsilon _{h,k}\right) \left( \omega +\varepsilon
_{h,k}\right) \beta _{h,k}\left( \varepsilon _{h,k}\right) d\varepsilon
_{h,k},  \label{Sigma11A}
\end{equation}
}
and the off-diagonal element is
{\small
\begin{equation}
\hat{\Sigma}_{21(12)}^{0}=-\frac{\Delta }{\pi }\text{sign(}j\text{)}%
\sum_{h}\int \Gamma _{h}\left( \varepsilon _{h,k}\right) \beta _{h,k}\left(
\omega \right) d\varepsilon _{h,k},  \label{Sigma21A}
\end{equation}
}
where the {\small$\beta _{h,k}\left( \omega \right) =\frac{g_{h,k}\left(
\omega \right) }{\left( \omega -\varepsilon _{h,k}\right) \left( \omega
+\varepsilon _{h,k}\right) g_{h,k}\left( \omega \right) -\Delta ^{2}}%
,g_{h,k}\left( \omega \right) =\frac{\omega ^{2}-\varepsilon _{-h,k}^{2}}{%
\omega ^{2}-\varepsilon _{k}^{2}},$} and the coupling {\small $\Gamma _{h}\left(
\varepsilon _{h,k}\right) =\Gamma _{0}\rho _{h}\left( \varepsilon
_{h,k}\right) $} with {\small $\Gamma _{0}=\pi \left\vert V\right\vert ^{2}/2D$}. The
density of states {\small $\rho _{h}\left( \varepsilon _{h,k}\right) =\frac{k}{4\pi
\left( d\varepsilon _{h,k}/dk\right) }$} can be explicitly evaluated
{\small
\begin{equation}
\rho _{h}\left( \varepsilon _{h,k}\right) =\left\{
\begin{array}{c}
\frac{m^{\star }}{2\pi \hbar ^{2}}\frac{\varepsilon _{R}}{\sqrt{\varepsilon
_{R}(\varepsilon _{k,h}-\varepsilon _{0})}}\delta _{h,-1}\text{ ;\ }%
E_{0}-\varepsilon _{R}<\varepsilon _{k,h}<E_{0}, \\
\frac{m^{\star }}{2\pi \hbar ^{2}}\left( \frac{1}{2}-\frac{h\varepsilon _{R}%
}{2\sqrt{\varepsilon _{R}\left( \varepsilon _{k,h}-\varepsilon _{0}\right) }}%
\right) \ \text{; \ }E_{0}<\varepsilon _{k,h}<D_{h}, \\
0\text{ \ \ \ \ \ \ \ \ otherwise,}%
\end{array}%
\right.  \label{Rho-app}
\end{equation}
}
where {\small $D_{h}=D+2h\sqrt{\left( D-E_{0}\right) \varepsilon _{R}}$}, {\small $D$} is the
half-band width, {\small $E_{0}$} is the bottom of the conduction band without
spin-orbit coupling, and {\small $\varepsilon _{R}=\frac{k_{0}^{2}}{2m^{\star }}$, $
k_{0}=\frac{m^{\star }\lambda _{R}}{\hslash }$}.

In general, it is difficult to treat the interacting self-energy exactly by the theoretical and
numerical approaches in all parameter regions. The off-diagonal element of {\small $%
\hat{F}_{d,j}\left( \omega \right)$} stands for the superconducting
correlations on impurity level, which is approximately given by the
Hartree-Fock approximation, like {\small $\langle \langle d_{-j}^{\dagger
}n_{j};d_{j}^{\dagger }\rangle \rangle =\langle n_{j}\rangle \langle \langle
d_{-j}^{\dagger };d_{j}^{\dagger }\rangle \rangle -\langle d_{j}^{\dagger
}d_{-j}^{\dagger }\rangle \langle \langle d_{j};d_{j}^{\dagger }\rangle
\rangle$} \cite{Cuevas2001,Vecino2003}.
The approximation is properly in large-U case, because the superconducting correlation
is significantly suppressed by the Coulomb repulsion.
From Eq.(\ref{DB2}), we obtain
{\small
\begin{equation}
\langle \langle d_{-j}^{\dagger };d_{j}^{\dagger }\rangle \rangle =\frac{
\hat{\Sigma}_{21}^{0}\left( \omega \right) +U\langle d_{j}^{\dagger
}d_{-j}^{\dagger }\rangle }{\omega +\varepsilon _{-j}+U\langle n_{j}\rangle
- \hat{\Sigma}_{11}^{0}\left( \omega \right) }\langle \langle
d_{j};d_{j}^{\dagger }\rangle \rangle .  \label{GFAN}
\end{equation}
}
Here, our aim is to obtain the interaction between the Kondo effect and
superconductivity. Therefore, we treat the off-diagonal elements of {\small $\hat{F}%
_{dj}\left( \omega \right)$} by Lacroix's approximation, which is believed to
properly capture the Kondo physics even at low temperatures.\cite%
{Luo1999,Lacroix1981} In the following, we show the main procedure to obtain
the diagonal elements of the Nambu GF.

From the Eq.(\ref{EM}), the EOM of the high order GF {\small $[\hat{F}_{d,j}(\omega
)]_{11}$} is
{\small
\begin{eqnarray}
&&\left( \omega -\varepsilon _{j}-U\right) \langle \langle
d_{j}n_{-j};d_{j}^{\dagger }\rangle \rangle  \notag \\
&=&\langle n_{-j}\rangle +\sum_{h}h^{-\left( j-1/2\right) }\int dk\tilde{V}%
_{k}\langle \langle \tilde{c}_{h,k,j}n_{-j};d_{j}^{\dagger }\rangle \rangle
\notag \\
&&+\sum_{h}h^{j+1/2}\int dk\tilde{V}_{k}\langle \langle d_{-j}^{\dagger }%
\tilde{c}_{h,k,-j}d_{j};d_{j}^{\dagger }\rangle \rangle  \notag \\
&&-\sum_{h}h^{j+1/2}\int dk\tilde{V}_{k}\langle \langle \tilde{c}%
_{h,k,-j}^{\dagger }d_{-j}d_{j};d_{j}^{\dagger }\rangle \rangle.
\label{Gdnd}
\end{eqnarray}
}
In above equation, the GF $\langle \langle d_{j}n_{-j};d_{j}^{\dagger }\rangle \rangle $
creates more higher order GFs, which can also be expanded by the EOM approach as following
{\small
\begin{eqnarray}
&&\left( \omega -\varepsilon _{h,k}\right) \langle \langle \tilde{c}%
_{h,k,j}n_{-j};d_{j}^{\dagger }\rangle \rangle  \notag \\
&=&h^{-\left( j-1/2\right) }\tilde{V}_{k}\langle \langle
d_{j}n_{-j};d_{j}^{\dagger }\rangle \rangle  \notag \\
&&+\sum_{h^{\prime }}h^{\prime \left( j+1/2\right) }\int dk^{\prime }\tilde{V%
}_{k^{\prime }}\langle \langle \tilde{c}_{h,k,j}d_{-j}^{\dagger }\tilde{c}%
_{h^{\prime },k^{\prime },-j};d_{j}^{\dagger }\rangle \rangle  \notag \\
&&-\sum_{h^{\prime }}h^{\prime \left( j+1/2\right) }\int dk^{\prime }\tilde{V%
}_{k^{\prime }}\langle \langle \tilde{c}_{h^{\prime },k^{\prime
},-j}^{\dagger }d_{-j}\tilde{c}_{h,k,j};d_{j}^{\dagger }\rangle \rangle
\notag \\
&&-\Delta \delta _{j,-1/2}\sum\limits_{h^{\prime }}h^{\prime }\langle
\langle \tilde{c}_{h^{\prime },-k,-j}^{\dagger }n_{-j};d_{j}^{\dagger
}\rangle \rangle  \notag \\
&&+\Delta \delta _{j,1/2}h\sum\limits_{h^{\prime }}\langle \langle \tilde{c}%
_{h^{\prime },-k,-j}^{\dagger }n_{-j};d_{j}^{\dagger }\rangle \rangle
\label{Gdcnd}
\end{eqnarray}
}
with $\varepsilon _{h,k}=\varepsilon _{k}+h\lambda _{R}k,$
{\small
\begin{eqnarray}
&&\left( \omega -\varepsilon _{h,k}+\varepsilon _{-j}-\varepsilon
_{j}\right) \langle \langle d_{-j}^{\dagger }\tilde{c}_{h,k,-j}d_{j};d_{j}^{%
\dagger }\rangle \rangle  \notag \\
&=&\langle d_{-j}^{\dagger }\tilde{c}_{h,k,-j}\rangle +h^{\left(
j+1/2\right) }\tilde{V}_{k}\langle \langle d_{j}n_{-j};d_{j}^{\dagger
}\rangle \rangle  \notag \\
&&-\sum_{h^{\prime }}h^{\prime \left( j+1/2\right) }\int dk^{\prime }\tilde{V
}_{k^{\prime }}\langle \langle \tilde{c}_{h^{\prime },k^{\prime
},-j}^{\dagger }\tilde{c}_{h,k,-j}d_{j};d_{j}^{\dagger }\rangle \rangle
\notag \\
&&+\sum_{h^{\prime }}h^{^{\prime }\left( -j+1/2\right) }\int dk^{\prime }%
\tilde{V}_{k^{\prime }}\langle \langle d_{-j}^{\dagger }\tilde{c}_{h,k,-j}%
\tilde{c}_{h^{\prime },k^{\prime },j};d_{j}^{\dagger }\rangle \rangle  \notag
\\
&&-\Delta \delta _{j,1/2}\sum\limits_{h^{\prime }}h^{\prime }\langle
\langle d_{-j}^{\dagger }\tilde{c}_{h^{\prime },-k,j}^{\dagger
}d_{j};d_{j}^{\dagger }\rangle \rangle  \notag \\
&&+\Delta \delta _{-j,1/2}h\sum\limits_{h^{\prime }}\langle \langle
d_{-j}^{\dagger }\tilde{c}_{h^{\prime },-k,j}^{\dagger }d_{j};d_{j}^{\dagger
}\rangle \rangle ,  \label{Gddcd}
\end{eqnarray}
}
and
{\small
\begin{eqnarray}
&&\left( \omega +\varepsilon _{h,k}-\varepsilon _{-j}-\varepsilon
_{j}-U\right) \langle \langle \tilde{c}_{h,k,-j}^{\dagger
}d_{-j}d_{j};d_{j}^{\dagger }\rangle \rangle  \notag \\
&=&\langle \tilde{c}_{h,k,-j}^{\dagger }d_{-j}\rangle -\tilde{V}%
_{k}h^{(j+1/2) }\langle \langle d_{j}n_{-j};d_{j}^{\dagger
}\rangle \rangle  \notag \\
&&+\sum_{h^{\prime }}h^{\prime \left( j+1/2\right) }\int dk^{\prime }%
\tilde{V}_{k^{\prime }}\langle \langle \tilde{c}_{h,k,-j}^{\dagger }\tilde{c}%
_{h^{\prime },k^{\prime },-j}d_{j};d_{j}^{\dagger }\rangle \rangle  \notag \\
&&+\sum_{h^{\prime }}h^{\prime \left( -j+1/2\right) }\int dk^{\prime }\tilde{%
V}_{k^{\prime }}\langle \langle \tilde{c}_{h,k,-j}^{\dagger }d_{-j}\tilde{c}%
_{h^{\prime },k^{\prime },j};d_{j}^{\dagger }\rangle \rangle  \notag \\
&&-\Delta \delta _{j,-1/2}h\sum\limits_{h^{\prime }}\langle \langle \tilde{c}%
_{h^{\prime },-k,j}d_{-j}d_{j};d_{j}^{\dagger }\rangle \rangle \notag \\
&&+\Delta \delta _{j,1/2}\sum\limits_{h^{\prime }}h^{\prime }\langle \langle
\tilde{c}_{h^{\prime },-k,j}d_{-j}d_{j};d_{j}^{\dagger }\rangle \rangle.
\label{Gdccd}
\end{eqnarray}
}
The Lacroix's approximation can be reached by taking the mean field
in the higher order GFs produced in Eq.(\ref%
{Gdcnd})-Eq.(\ref{Gdccd}), such as $\langle \langle \tilde{c}%
_{h,k,j}d_{-j}^{\dagger }\tilde{c}_{h^{\prime },k^{\prime
},-j};d_{j}^{\dagger }\rangle \rangle \approx \langle d_{-j}^{\dagger }\tilde{c}%
_{h^{\prime },k^{\prime },-j}\rangle \langle \langle \tilde{c}%
_{h,k,j};d_{j}^{\dagger }\rangle \rangle +\langle \tilde{c}_{h^{\prime
},k^{\prime },-j}\tilde{c}_{h,k,j}\rangle \langle \langle d_{-j}^{\dagger
};d_{j}^{\dagger }\rangle \rangle $, where the second term involving
the superconducting correlations ({\small0($\Delta^2$)}) can be neglected.
Furthermore, we neglect the higher GFs
containing the superconducting correlation on the impurity due to {\small $U\gg
\Delta $}, such as {\small $\langle \langle \tilde{c}_{h^{\prime },-k,-j}^{\dagger
}n_{-j};d_{j}^{\dagger }\rangle \rangle \ $} and {\small $\langle \langle \tilde{c}%
_{h^{\prime },-k,j}d_{-j}d_{j};d_{j}^{\dagger }\rangle \rangle $}.
After some straightforward algebraic operations, we gain the higher order GF
{\small
\begin{eqnarray}
&&\langle \langle d_{j}n_{-j};d_{j}^{\dagger }\rangle \rangle =O_{j}\left( \omega \right)
+Q_{j}\left( \omega \right)\langle \langle d_{j};d_{j}^{\dagger }\rangle \rangle
+P_{j}\left( \omega \right)\langle \langle d_{-j}^{\dagger
};d_{j}^{\dagger }\rangle \rangle , \notag \\
&&  \label{Gnddf}
\end{eqnarray}
}
Then, we obtain the GF
{\small
\begin{equation}
\langle \langle d_{j};d_{j}^{\dagger }\rangle \rangle =\frac{1+UO_{j}\left(
\omega \right) +\left( 1+UP_{j}\left( \omega \right) \right) \hat{\Sigma}_{21}^{0}\left( \omega \right)
\langle \langle d_{-j}^{\dagger };d_{j}^{\dagger }\rangle \rangle\left(
\omega \right) }{\omega -\varepsilon _{j}-\hat{\Sigma}_{11}^{0}\left( \omega
\right) -UQ_{j}\left( \omega \right) },  \label{GFD}
\end{equation}
}
by substituting Eq.(\ref{Gnddf}) into Eq.({\ref{GmatrixApp}}).
The notations introduced are
{\small
\begin{equation}
O_{j}\left( \omega \right) =\frac{\langle n_{-j}\rangle +A_{1,j}\left(
\omega \right) -A_{2,j}\left( \omega \right) }{\omega -\varepsilon _{j}-U-\Xi _{0}\left( \omega \right) -\Xi _{1}\left(
\omega \right) -\Xi _{2}\left( \omega \right) },  \label{Oj}
\end{equation}
}
{\small
\begin{equation}
P_{j}\left( \omega \right) =\frac{\left( A_{1,j}\left( \omega \right)
-A_{2,j}\left( \omega \right) \right) \hat{\Sigma}_{21}^{0}\left( \omega
\right) }{\omega -\varepsilon _{j}-U-\Xi _{0}\left( \omega \right) -\Xi _{1}\left(
\omega \right) -\Xi _{2}\left( \omega \right) },  \label{Pj}
\end{equation}
}
{\small
\begin{equation}
Q_{j}\left( \omega \right) =\frac{\left( A_{1,j}\left( \omega \right)
-A_{2,j}\left( \omega \right) \right) \hat{\Sigma}_{11}^{0}\left( \omega
\right) -\left( B_{1,j}\left( \omega \right) +B_{2,j}\left( \omega \right)
\right) }{\omega -\varepsilon _{j}-U-\Xi _{0}\left( \omega \right) -\Xi _{1}\left(
\omega \right) -\Xi _{2}\left( \omega \right) } . \label{Qj}
\end{equation}
}
where {\small $\Xi _{0}\left( \omega \right) =\sum_{h}\int dk\frac{\left\vert \tilde{%
V}_{k}\right\vert ^{2}}{\omega -\varepsilon _{h,k}}$, $\Xi _{1}\left( \omega
\right) =\sum_{h}\int dk\frac{\left\vert \tilde{V}_{k}\right\vert ^{2}}{%
\omega -\varepsilon _{1,j,h,k}}$, $\Xi
_{2}\left( \omega \right) =\sum_{h}\int dk\frac{\left\vert \tilde{V}%
_{k}\right\vert ^{2}}{\omega +\varepsilon _{2,j,h,k}}$}
with {\small $\varepsilon _{1,j,h,k}=-\varepsilon _{h,k}+\varepsilon
_{-j}-\varepsilon _{j}$} and {\small $\varepsilon _{2,j,h,k}=\varepsilon
_{h,k}-\varepsilon _{-j}-\varepsilon _{j}-U$}.
By transfer the summation of $k$ into integration, the notations
{\small $A_{\eta ,j}\left( \omega \right) $} and {\small $B_{\eta ,j}\left( \omega \right)
$ ($\eta =1,2$)} can be obtained by the spectral theorem.
Taking some simplification procedures, we obtain
{\small
\begin{equation}
A_{\eta ,j}\left( \omega \right) =\frac{i}{2\pi ^{2}}\sum_{h}\int
d\varepsilon _{h,k}\frac{\Gamma _{h}\left( \varepsilon _{h,k}\right) \Theta
_{h}\left( \varepsilon _{h,k}\right) }{\omega -\varepsilon _{\eta ,j,h,k}}
\label{Aej}
\end{equation}
}
and
{\small
\begin{equation}
B_{\eta ,j}\left( \omega \right) =\frac{i}{\pi ^{2}}\sum_{h}\int
d\varepsilon _{h,k}\frac{\Gamma _{h}\left( \varepsilon _{h,k}\right) \Xi
_{h}\left( \varepsilon _{h,k}\right) }{\omega -\varepsilon _{\eta ,j,h,k}},
\label{Bej}
\end{equation}
}
where {\small $\Theta _{h}\left( \varepsilon _{h,k}\right) =\int f\left( \omega
\right) [\left( \omega +\varepsilon _{h,k}\right) \beta _{h,k}\left( \omega
\right) \hat{G}_{d,-j}\left( \omega \right) _{11}-c.c]d\omega $} and {\small $\Xi
_{h}\left( \varepsilon _{h,k}\right) =\int f\left( \omega \right) \left(
\omega +\varepsilon _{h,k}\right) \beta _{h,k}\left( \omega \right) d\omega$}.

\end{document}